\pdfoutput=1 

\documentclass[11pt,letterpaper,twocolumn]{scrartcl}

\makeatletter
\DeclareOldFontCommand{\rm}{\normalfont\rmfamily}{\mathrm}
\DeclareOldFontCommand{\sf}{\normalfont\sffamily}{\mathsf}
\DeclareOldFontCommand{\tt}{\normalfont\ttfamily}{\mathtt}
\DeclareOldFontCommand{\bf}{\normalfont\bfseries}{\mathbf}
\DeclareOldFontCommand{\it}{\normalfont\itshape}{\mathit}
\DeclareOldFontCommand{\sl}{\normalfont\slshape}{\@nomath\sl}
\DeclareOldFontCommand{\sc}{\normalfont\scshape}{\@nomath\sc}
\makeatother

\usepackage[top=2.4cm, bottom=2.4cm, left=1.5cm, right=1.5cm,footskip=0.8cm]{geometry}

\usepackage[T1]{fontenc}
\usepackage[utf8]{inputenc}
\usepackage{abstract}

\usepackage[hidelinks]{hyperref}
\usepackage[parfill]{parskip}
\usepackage[separate-uncertainty = true,multi-part-units=single]{siunitx}
\usepackage{slashed}
\usepackage{microtype}
\usepackage{csquotes}
\usepackage{graphicx}
\usepackage{amsmath}
\usepackage{amssymb}
\usepackage{braket}
\usepackage{booktabs}
\usepackage{caption}
\usepackage{subcaption}
\usepackage{todonotes}
\usepackage{grffile}
\usepackage{enumitem}

\usepackage[numbers,sort&compress]{natbib}

\usepackage{multirow}

\usepackage{fancyhdr}
\usepackage[yyyymmdd,hhmmss]{datetime}
\fancypagestyle{plain}
{
}
\usepackage{cleveref}

\usepackage[auth-sc,affil-it]{authblk}

\usepackage{scalefnt}
\newcommand{\abbrev}{\scalefont{.9}}

\newcommand{\NNLL}{\text{\abbrev N$^2$LL}}
\newcommand{\NNNLL}{\text{\abbrev N$^3$LL}}
\newcommand{\NNNLLp}{\text{\abbrev N$^3$LL'}}
\newcommand{\NFLL}{\text{\abbrev N$^4$LL}}
\newcommand{\NFLLp}{\text{\abbrev N$^4$LL$_\text{p}$}}
\newcommand{\NNLO}{\text{\abbrev NNLO}}
\newcommand{\NNNLO}{\text{\abbrev N$^3$LO}}
\newcommand{\NLO}{\text{\abbrev NLO}}

\newcommand{\EFT}{\text{\abbrev EFT}}

\newcommand{\HPC}{\text{\abbrev HPC}}
\newcommand{\SM}{\text{\abbrev SM}}

\newcommand{\IR}{\text{\abbrev IR}}

\newcommand{\RG}{\text{\abbrev RG}}

\newcommand{\QCD}{\text{\abbrev QCD}}
\newcommand{\EW}{\text{\abbrev EW}}

\newcommand{\PDF}{\text{\abbrev PDF}}

\newcommand{\LHC}{\text{\abbrev LHC}}

\newcommand{\CMS}{\text{\abbrev CMS}}
\newcommand{\ATLAS}{\text{\abbrev ATLAS}}

\newcommand{\SCET}{\text{\abbrev SCET}}
\newcommand{\RGE}{\text{\abbrev RGE}}

\newcommand{\MCFM}{\text{\abbrev MCFM}}

\newcommand{\CuTeMCFM}{\texttt{CuTe-MCFM}}

\newcommand{\abs}[1]{\lvert#1\rvert}

\newcounter{notecount}

\makeatletter
\renewcommand\maketitle{
	\begin{center}
		{\huge\bfseries\@title\par\vspace{0.3em}}
		{\scshape\@author, \@date}
	\end{center}
}
\makeatother

\fancypagestyle{firstpage}{%
	\lhead{}
	\rhead{FERMILAB-PUB-22-528-T}
}

\begin{document}
	
	\thispagestyle{firstpage}
	\title{\Large { Fiducial Drell-Yan production at the LHC improved by transverse-momentum 
	resummation at N$^4$LL$_\text{p}$+N$^3$LO }}
	
	\author[1]{Tobias Neumann}
	\author[2]{John Campbell}

	\affil[1]{Department of Physics, Brookhaven National Laboratory, Upton, New York 11973, USA}
	\affil[2]{Fermilab, PO Box 500, Batavia, Illinois 60510, USA}

	\date{}
	\twocolumn[
	\maketitle
	
	\vspace{0.5cm}
	
	\begin{onecolabstract}
		\vspace{0.5cm} 
		
		Drell-Yan production is one of the precision cornerstones of the \LHC{}, serving as 
		calibration for measurements such as the $W$-boson mass. Its extreme precision at the level 
		of 
		1\% challenges theory predictions at the highest level. We present the first independent 
		calculation of Drell-Yan production at order $\alpha_s^3$ in transverse-momentum ($q_T$) 
		resummation 
		improved perturbation theory. Our calculation reaches the state-of-the-art through 
		inclusion of the recently 
		published four loop rapidity anomalous dimension and three loop massive 
		axial-vector contributions. We compare to the most recent data from \CMS{} with fiducial 
		and 
		differential cross-section predictions and find excellent agreement at the percent level. 
		Our resummed calculation including the matching to 
		$Z$+jet 
		production at \NNLO{} is publicly available in the upcoming \CuTeMCFM{} 10.3 release and 
		allows for theory-data 
		comparison at an unprecedented level.
		\vspace{0.5cm}
	\end{onecolabstract}
	]
	
	\vspace{-5mm}
	
	\tableofcontents

	\section{Introduction}
	
	Drell-Yan ($Z$-boson) production is among the most important standard candles of the high-energy 
	\LHC{} 
	physics program due to its very precise measurement at the level of one percent 
	\cite{ATLAS:2015iiu,CMS:2016mwa,CMS:2019raw,ATLAS:2019zci}. 
	It is used for the 
	extraction 
	of the strong 
	coupling \cite{Ball:2018iqk,Camarda:2022qdg}, fitting of parton distribution functions 
	\cite{Boughezal:2017nla,Giuli:2020bkp} that further constrain and determine 
	Standard Model (\SM{}) input parameters, and is also a crucial ingredient of
	the $W$-boson mass determination 
	\cite{ATLAS:2017rzl,LHCb:2021bjt,CDF:2022hxs}.
	
	The current precision in \QCD{} for Drell-Yan predictions is at the level of $\alpha_s^3$
	both fully differentially \cite{Bizon:2019zgf,Camarda:2021ict,Re:2021con,Chen:2022cgv} and more 
	inclusively \cite{Duhr:2021vwj,Chen:2021vtu}.
	Calculations at this order have been performed at fixed order (\NNNLO{}) and including
	the effects of transverse momentum ($q_T$) resummation up to \NNNLL{} logarithmic accuracy. 
	Currently all fully differential calculations at the level of $\alpha_s^3$ employ transverse 
	momentum
	subtractions or 
	transverse momentum resummation. They have been enabled by the recent availability of the 
	three-loop 
	beam-functions \cite{Luo:2020epw,Ebert:2020yqt,Luo:2019szz}, complete three-loop hard function 
	\cite{Gehrmann:2010ue,Baikov:2009bg,Lee:2010cga,Gehrmann:2021ahy,Chen:2021rft} and the existence
	of a \NNLO{} calculation of $Z$+jet production
	\cite{Boughezal:2015dva,Gehrmann-DeRidder:2015wbt,Boughezal:2015ded,Gehrmann-DeRidder:2017mvr,Boughezal:2016dtm}.
	Beyond pure QCD corrections, the full set of two-loop mixed 
	\QCD{}$\otimes$\EW{} corrections 
	have been calculated very recently \cite{Heller:2020owb,Bonciani:2021zzf,Buccioni:2022kgy}.
	
	Traditionally there has been a focus on fixed-order calculations for total 
	fiducial cross-sections, but now that relatively high perturbative 
	orders have been reached, convergence issues of the 
	perturbative series due to fiducial cuts have been identified
        \cite{Ebert:2020dfc,Salam:2021tbm,Billis:2021ecs}. These issues trace 
back to a linear sensitivity of acceptance cuts to small transverse momenta, where fixed-order 
predictions are unreliable, leading to factorially divergent contributions \cite{Salam:2021tbm}. It has shifted 
the focus towards resummation-improved results even for total fiducial cross-sections, which can
cure such problems without requiring any modification of analysis cuts.

All calculations matched to \NNLO{} $Z$+jet fixed-order at large $q_T$ have so far been 
based on the {\abbrev NNLOjet} results \cite{Gehrmann-DeRidder:2015wbt}. Different 
implementations of $q_T$ resummation and subtractions are built on top of this calculation. 
Results for a matching 
to the resummation in {\abbrev DYTurbo} \cite{Camarda:2019zyx} have been presented in 
ref.~\cite{Camarda:2021ict} where 
only non-singlet and vector singlet\footnote{
In singlet contributions the $Z$ boson does not directly couple to 
the incoming quarks, but is separated through loops involving gluons. These contributions therefore 
only enter at higher orders.}
contributions are included and truncation uncertainties are estimated 
by 
considering differences between successive orders.
 A matching to the 
{\abbrev RadISH} 
resummation approach \cite{Re:2021con,Bizon:2017rah} has been presented in 
refs.~\cite{Bizon:2019zgf,Re:2021con}, also neglecting axial singlet 
contributions. Axial singlet contributions in the $m_t\to\infty$ \EFT{} have been included in the 
resummed calculation of ref.~\cite{Ju:2021lah} but without the matching to $\alpha_s^3$ 
fixed-order. The {\abbrev NNLOjet} 
setup has subsequently been extended to calculate 
fiducial cross-sections also at fixed-order \NNNLO{}, comparing the impact of power 
corrections through studying the difference between symmetric and product cuts \cite{Chen:2022cgv} 
and comparing with \SI{13}{\TeV} \ATLAS{} data  \cite{ATLAS:2019zci}. The {\abbrev RadISH} based 
calculations provide uncertainty estimates for differential and fiducial results for the first 
time. 

 Despite these studies, it is crucial to 
have an independent calculation of both the fixed-order components and the 
resummation implementation.
While the {\abbrev NNLOjet} calculation is tested by the correct approach of the triple 
singular limits through an implementation of 
(differential) $q_T$ subtractions, it is important to also probe the finite contributions.
As well as acting as a cross-check, an additional calculation also provides
an independent estimate of uncertainties.
		
		In this paper we present both a publicly available calculation of $Z$-boson production as 
		well 
		as differential and fiducial cross-sections at the state-of-the-art level 
		\NFLLp{}+\NNNLO{}. The \enquote{p} subscript denotes that we are $\alpha_s^3$ accurate in 
		fixed-order and \RG{}-improved perturbation theory up to missing effects from exact 
		\NNNLO{} \PDF{}s that contribute both to fixed-order and logarithmic accuracy at 
		$\alpha_s^3$. We 
		include the four loop rapidity anomalous 
		dimension \cite{Duhr:2022yyp,Moult:2022xzt}, pushing the accuracy to this level for the 
		first time. We also include the 
		massive three-loop axial singlet contributions \cite{Chen:2021rft} without 
		the need for approximations.
		We compare at $\alpha_s^3$ accuracy with the \CMS{} 
		\SI{13}{\TeV} precision measurement. All parts, both resummation and fixed-order are 
		publicly 
		available in 
		the 
		next 
		\CuTeMCFM{} release 10.3. Public codes are crucial to ensure reproducibility, allow the 
		community to perform independent checks, to calculate predictions with different 
		parameters, and provide the 
		basis for future theoretical
		improvements as strongly advocated by our community \cite{Caola:2022ayt}. 
		
	In \cref{sec:calculation} we provide technical details of our calculation before presenting 
	results in \cref{sec:results} and concluding in \cref{sec:conclusions} with an outlook.
	
	\section{Calculation}
	\label{sec:calculation}

	We consider \QCD{} corrections to the process $q+\bar{q} \to Z/\gamma (\to 
	l^-+l^+)$.
	Our calculation in \CuTeMCFM{} \cite{Becher:2020ugp,Campbell:2019dru}  matches resummation at 
	the level of \NFLLp{} to $\alpha_s^3$ fixed-order $Z$+jet production.
	 Apart from missing 
	\NNNLO{} \PDF{} effects we achieve full $\alpha_s^3$ fixed-order and transverse momentum 
	renormalization-group-improved (\RG{}-improved) logarithmic accuracy by counting 
	$\log(q_T^2/Q^2)\sim1/\alpha_s$.\footnote{While 
	we are neglecting \NNNLO{} \PDF{}s for full 
	\NFLL{}+\NNNLO{} accuracy, it has been customary in the literature to refer to predictions as 
	\NNNLO{} despite the lack of these corrections.
	 } We further calculate fixed-order \NNNLO{} results based on $q_T$-subtractions.
        Our calculation involves many contributions at the fixed-order and 
        at the resummation level, which we discuss separately below.

	\subsection{Resummation}
 The resummation is based on the \SCET{} formalism derived in 
refs.~\cite{Becher:2010tm,Becher:2011xn,Becher:2012yn} and originally implemented as 
\CuTeMCFM{} in ref.~\cite{Becher:2020ugp} to \NNNLL{}. Large logarithms $\log(q_T^2/Q^2)$ are 
resummed through \RG{} evolution of hard- and beam functions in a small-$q_T$ factorization 
theorem. Rapidity logarithms are directly exponentiated through the collinear-anomaly 
formalism.	
	
At large $q_T$ the small-$q_T$ factorization theorem becomes invalid and one has 
to switch 
to fixed-order predictions. We switch using a transition function that smoothly 
interpolates between resummation and fixed-order without disturbing subleading power 
corrections, 
as detailed in ref.~\cite{Becher:2020ugp}. Within this procedure the overlap between 
fixed-order and resummation has to be subtracted by expanding the resummation to a fixed-order.  
This difference is referred to as matching corrections. For $Z$ boson production they quickly 
approach zero for $q_T\to0$ and remain at the few percent level up to $\sim\SI{30}{\GeV}$, see our 
dedicated discussion below.	
	
	Three loop transverse momentum dependent beam functions have been calculated in refs.~
	\cite{Luo:2020epw,Ebert:2020yqt,Luo:2019szz} and implemented in ref.~\cite{Neumann:2021zkb} in 
	\CuTeMCFM{}. Together with the $\alpha_s^3$ hard function this enables resummation at the level 
	of \NNNLLp{}.
	  The resummation of linear 
	power corrections~\cite{Ebert:2020dfc} has been included in \CuTeMCFM{} since its initial 
	implementation through a recoil prescription~\cite{Becher:2019bnm}. They are crucial to improve 
	the resummation itself as well as the numerical stability by allowing a larger matching cutoff
	(the value of $q_T$ below which matching corrections are set to zero). It is also crucial for 
	the stability of our fixed-order \NNNLO{} results in the presence of symmetric lepton cuts, see 
	below.
	
	In this study we have upgraded the resummation to the logarithmic accuracy of \NFLLp{} through 
	the inclusion of the four loop rapidity anomalous dimension \cite{Duhr:2022yyp,Moult:2022xzt}. 
	 While the five 
	loop 
	cusp anomalous dimension is also a necessary 
	ingredient, it only enters through the hard function evolution and is numerically completely 
	negligible. Already at a lower order the hard function evolution is precise at the level of one 
	per-mille. We nevertheless include such effects in the hard function evolution by taking four 
	loop 
	collinear 
	anomalous dimensions from ref.~\cite{Agarwal:2021zft} and a five loop cusp estimate from 
	ref.~\cite{Herzog:2018kwj} that agrees with our own Pad\'e approximant estimate. The five loop 
	beta function is taken from ref.~\cite{Baikov:2016tgj}.

	Transverse momentum Fourier conjugate logarithms $L_\perp \sim \log(x_T^2 \mu^2)$ appearing in 
	the factorization theorem would traditionally be integrated over the full range of $x_T$.
	This requires the introduction of a prescription to avoid the Landau pole. Following the 
	\SCET{} resummation 
	formalism 
	of ref.~\cite{Becher:2010tm,Becher:2011xn} this is not necessary as scales are always set in 
	the perturbative regime. The formalism further employs an improved power counting $L_\perp 
	\sim 
	1/\sqrt{\alpha_s}$ that is crucial to improve the resummation at small $q_T$ 
	\cite{Becher:2011xn}. At \NFLL{} the $\alpha_s^3$ beamfunctions 
	\cite{Luo:2020epw,Ebert:2020yqt,Luo:2019szz} are then not sufficient for \emph{improved}
	$\alpha_s^3$ accuracy. Using the beamfunction \RGE{}s we reconstructed the logarithmic 
	beamfunction terms up to order $\alpha_s^6 L_\perp^6$, $\alpha_s^4 
	L_\perp^4$ and $\alpha_s^4 L_\perp^2$. We performed the Mellin convolutions of beam function 
	kernels and splitting functions up to three loops \cite{Vogt:2004mw,Moch:2004pa} using the 
	{\abbrev MT} package \cite{Hoschele:2013pvt}.

The hard function entering the factorization formula consists of {\abbrev 
$\overline{\text{MS}}$}-renormalized virtual 
corrections. For Drell-Yan production one typically distinguishes between different classes of 
corrections based on 
the following decomposition.
The Feynman rule vertex for the photon coupling to fermions is $-ie Q_f \gamma^\mu$, while the 
Z coupling is $-ie\gamma^\mu(v_L^f P_L + v_R^f P_R)$. In terms of vector and axial-vector 
components this decomposes as 
\begin{gather}
	(v_L^f P_L + v_R^f P_R) = \left(\frac{1}{2} v_L^f + \frac{1}{2}v_R^f\right) - \gamma_5 
	\left(\frac{1}{2} v_L^f - \frac{1}{2} v_R^f\right)\,.
\end{gather}
The first term constitutes the vector coupling and is dressed by a vector form-factor 
$F_V$ that encapsulates higher-order corrections. The second term constitutes the axial-vector 
coupling and is dressed by an axial-vector form-factor $F_A$.
For a photon exchange $v_L=v_R=1$ and $F_A=0$. On the other hand, the coupling of $Z$ bosons to 
quarks involves both a vector ($F_V$) and an axial-vector ($F_A$)
form factor. 
A common 
approximation is to include only non-singlet contributions, which leads to $F_A=F_V$.

 The three-loop corrections to the vector 
	part have been known for a while now \cite{Gehrmann:2010ue,Baikov:2009bg,Lee:2010cga}, 
	while the three-loop corrections to the axial singlet part have only been 
	computed recently in purely massless \QCD{} \cite{Gehrmann:2021ahy} and with full top-quark 
	mass dependence \cite{Chen:2021rft}. In our calculation we include the complete three-loop 
	corrections with full top-quark mass dependence. While these contributions are small, the 
	top-quark mass dependence does not decouple in either the $m_t\to\infty$ limit or 
	the low-energy limit, in contrast to the vector case.
	
	\subsection{$Z$+jet \NNLO{} fixed order}
	Our fixed-order \NNLO{} $Z$+jet calculation is based on ref.~\cite{Boughezal:2015ded}, employing 
	1-jettiness subtractions \cite{Gaunt:2015pea,Boughezal:2015dva,Stewart:2010tn}. For 1-jettiness 
	subtractions at \NNLO{} a crucial new	ingredient compared to 
	0-jettiness is the \NNLO{} soft function which has been calculated in 
	refs.~\cite{Campbell:2017hsw,Boughezal:2015eha}. Top-quark loop corrections to $Z$+jet and 
	$Z$+2~jet production have been known analytically for 
	some time \cite{Campbell:2016tcu} and are included in our calculation. Two-loop axial singlet 
	contributions in the $Z$+jet hard function are unknown so far and have been neglected in our 
	calculation.

	We have performed extensive cross-checks of all elements of the calculation.
        We find numerical agreement between all bare amplitude expressions and 
	Recola \cite{Denner:2017wsf}, and have reproduced the non-singlet hard function that was 
	originally taken from the code {\abbrev PeTeR} \cite{Becher:2011fc,Becher:2013vva}
	with an independent re-implementation from 
	refs.~\cite{Gehrmann:2011ab,Garland:2002ak,Gehrmann:2002zr}.
	We have thoroughly tested the implementation of the subtraction terms using the
	same methodology as in ref.~\cite{Campbell:2019gmd}.  Compared to the original
	implementation \cite{Boughezal:2015ded} we identified an inconsistency in a small
	number of subtraction terms and in the crossing of one-loop axial-vector helicity
	amplitudes.  As a final check, we  compared with fiducial 
	results presented in ref.~\cite{Gehrmann-DeRidder:2016ycd} for different partonic channels and 
	find agreement.	
	
	\subsection{Matching corrections and cutoff effects}
	Since our calculation is based on $1$-jettiness slicing subtractions, unlike the local antenna 
	subtractions used in the {\abbrev NNLOjet} calculation \cite{Gehrmann-DeRidder:2015wbt}, we 
	have to pay attention to 
	residual slicing cutoff effects. 
	Jettiness slicing at the level of \NNLO{} in association with one jet is widely believed to 
	have reached its limits of applicability.  But, as we demonstrate in this paper, optimized 
	phase-space generation and the inclusion of linear power corrections together with an efficient 
	parallelization for the use of modern \HPC{} 
	resources \cite{Cordero:2022gsh} allows us to compute results at the level of 
	\NFLLp{}+\NNNLO{} and \NNNLO{} with negligible systematic cutoff uncertainties.
	
\begin{figure}
		\includegraphics[width=\columnwidth]{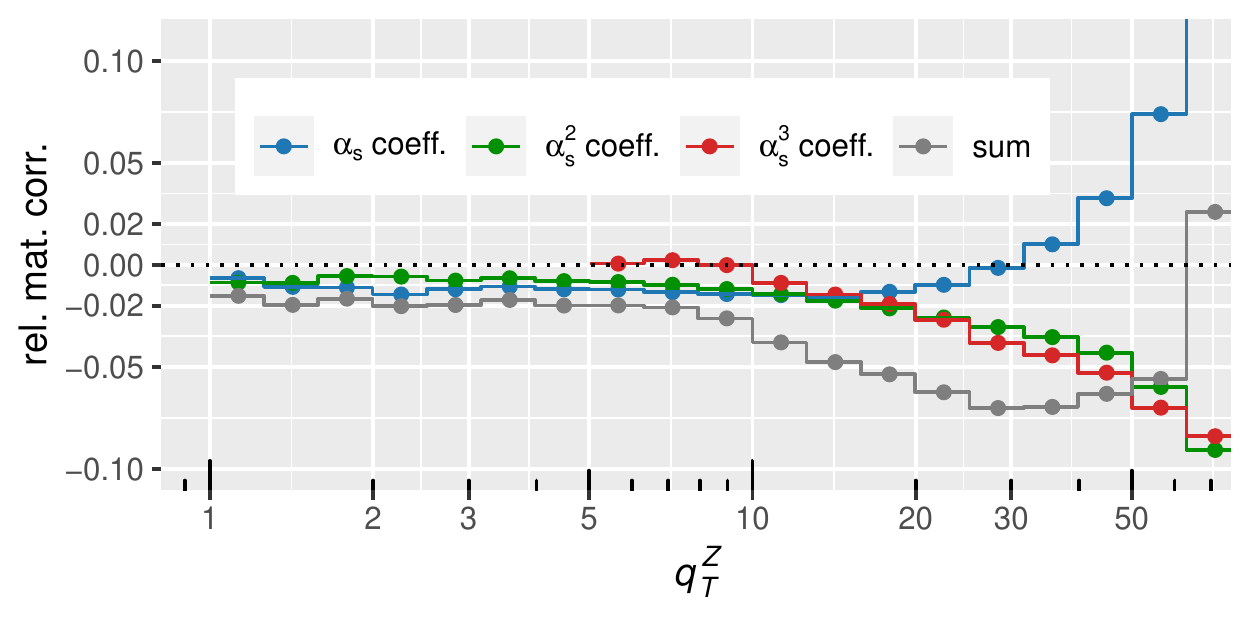}
		\caption{Matching corrections of the $\alpha_s,\alpha_s^2$ and $\alpha_s^3$ coefficients 
			relative to the naively matched result at \NFLL{} (matching 
			corrections + resummed result without transition function) for the \CMS{} cuts as in 
			the main 
			document.}
		\label{fig:matchcorr}
	\end{figure}

	  Nevertheless, we had to choose the $q_T$ cutoff for the resummation matching 
	corrections and the $q_T$ cutoff for the $q_T$ subtractions at \NNNLO{} low enough that 
	residual cutoff effects can be neglected:
		
	 In fig.~\ref{fig:matchcorr} we show the matching corrections of the $\alpha_s,\alpha_s^2$ and 
	$\alpha_s^3$ coefficients relative to the naively matched result at \NFLLp{} for the \CMS{} 
	analysis in the results section. The naively matched 
	result consists of matching corrections and resummed result without transition function. The 
	size 
	of the matching corrections on the one hand indicates where the transition function needs to 
	switch 
	between resummed and fixed-order calculations. In this case matching corrections become sizable 
	around \SI{50}{\GeV} and the resummation quickly breaks down beyond \SI{60}{\GeV}. This 
	motivates 
	our choice to use a transition function as detailed in 
	ref.~\cite{Becher:2020ugp} using $x_T^{\text max} = (q_T^{\text max}/M_Z)^2$
	with $q_T^{\text max}$ in the range $40$ to \SI{60}{\GeV}. The transition uncertainties are then
	comparable to uncertainties in the fixed-order and resummation region and we are therefore 
	minimally
	sensitive to the precise range and 
	shape of the transition.  

	At $\alpha_s$ and $\alpha_s^2$ matching corrections can be 
	safely neglected
	 below \SI{1}{\GeV}, but the numerical 
	implementation allows for smaller cutoffs if necessary.
	Figure~\ref{fig:matchcorr} further justifies our neglect of matching corrections below 
	\SI{5}{\GeV} at $\alpha_s^3$. The approach to zero of the matching corrections towards smaller 
	$q_T$ shows that 
	the large 
	logarithms present in the fixed-order and expanded resummation calculations cancel. While the 
	$\alpha_s^3$ matching corrections at 
	\SI{5}{\GeV} are zero within numerical uncertainty, from the lower order results we see 
	fluctuations
	at the level of one percent for smaller values of $q_T$. On the fiducial cross-section 
	we 
	therefore estimate an uncertainty due to missing matching corrections by multiplying the 
	resummed 
	result integrated up to \SI{5}{\GeV} with one percent. This is about \SI{1}{pb}, our quoted 
	numerical precision. Similarly the effect on the Z-boson $q_T$ distributions below \SI{5}{\GeV} 
	is 
	expected to be less than 1\%. This is also the region with substantial resummation 
	uncertainties 
	from a variation of the low scale. The effect is therefore negligible.  The size of the 
	corrections is in line 
	with the 
	findings of previous studies \cite{Chen:2022cgv,Camarda:2021ict}.

	The $\alpha_s^3$ coefficient of fig.~\ref{fig:matchcorr} has been obtained using a dynamic 
	$\tau_1^\text{cut}$ value of $7.6\cdot10^{-5}\cdot\sqrt{(q_T^Z)^2+m_Z^2}$, which is about 
	\SI{0.007}{\GeV} for small $q_T$. Our one-jettiness is defined by 
	\begin{gather}
		\tau_1 = \sum_{\text{partons}\, k} \min_i \left \{ \frac{2 r_i q_k}{Q_i} \right\}\,,
	\end{gather}
	where the sum over $i$ is over the two beam momenta and the jet axis determined by anti-$k_T$ 
	$R=0.5$
	clustering and $Q_i$ are chosen to be $2E_i$. We have checked the $\tau_1^\text{cut}$ 
	dependence to 
	determine that with the given $\tau_1^\text{cut}$ cutoff we can only reliably use a $q_T$ 
	resummation matching 
	cutoff of \SI{5}{\GeV}, as shown in fig.~\ref{fig:matchcorr}.
	
		Smaller matching cutoffs would require smaller 
	$\tau_1^\text{cut}$ values for the large $q_T$ logarithms to cancel between fixed-order \NNLO{} 
	Z+jet calculation and fixed-order expansion of the resummation, increasing computational costs 
	significantly:
For a $q_T$ cutoff of \SI{5}{\GeV} 
the small size of the $1$-jettiness parameter requires computing resources of about 6000 
{\abbrev 
	NERSC} Perlmutter node	hours for 
all fiducial and differential results presented in the following (we ran with 256 nodes for 
about one day). While a cutoff of 
\SIrange{2}{3}{\GeV} could likely be achieved with more resources (due to 
requiring a 
smaller jettiness parameter), the inclusion of subleading $1$-jettiness power corrections,
which have currently only been computed at a lower order \cite{Boughezal:2019ggi},
could be a more promising resource-saving approach.

	\subsection{\NNNLO{} fixed order}
	
	For our \NNNLO{} fixed-order results we have integrated the $q_T$ factorization theorem 
	expanded to 
	$\alpha_s^3$ over $q_T$ up to a slicing cutoff $q_T^\text{cut}$. This allows us to implement 
	$q_T$ subtractions by combining this contribution with the fixed-order $Z$+jet \NNLO{} 
	calculation, regulating 
	\IR{} divergences through the cutoff and extrapolating $q_T^\text{cut}$ to zero. We have 
	checked 
	that \NNLO{} results obtained with this implementation agree with previous implementations of 
	jettiness subtractions and $q_T$-subtractions in \MCFM{} 
	\cite{Boughezal:2016wmq,Campbell:2022gdq}.
	
	\begin{table}
		\centering
		\caption{Fiducial cuts for $Z\to l^+ l^-$ 
			used in the \ATLAS{} \SI{13}{\TeV} analysis \cite{ATLAS:2019zci}.}
		\vspace*{0.5em}
		\bgroup
		\setlength\tabcolsep{1em}
		\def\arraystretch{1.5}%
		\begin{tabular}{l|c}
			Lepton cuts & $q_T^{l} > \SI{27}{\GeV}, \abs{\eta^l} < 2.5$\\
			Mass cuts  & $\SI{66.0}{GeV} < m^{l^+l^-} < \SI{116.0}{GeV}$
		\end{tabular}
		\egroup
		\label{tab:zatlas13tev}
	\end{table}

	As an additional cross-check of our calculation	and validation of results in the literature, we 
	compare with the 
	fiducial cross-section results in 
	ref.~\cite{Chen:2022cgv} for the most challenging case of symmetric lepton cuts. The authors 
	employ 
	cuts 
	for 
	the \SI{13}{\TeV} \ATLAS{} analysis \cite{ATLAS:2019zci} as in 
	table~\ref{tab:zatlas13tev}.
	We furthermore adopt their choice of \PDF{}, {\texttt {NNPDF4.0}} at \NNLO{} with 
	$\alpha_s(m_Z)=0.118$ \cite{NNPDF:2021njg}, and the $G_\mu$ scheme with 
	$m_Z=\SI{91.1876}{\GeV}$, 
	$m_W=\SI{80.379}{\GeV}$, $\Gamma_Z=\SI{2.4952}{\GeV}$, $\Gamma_W=\SI{2.085}{\GeV}$, 
	$G_F=\SI{1.663787e-5}{\GeV^{-2}}$. 
	
	We can naturally fully reproduce their fixed-order results up to 
	\NNLO{}. At \NNNLO{} we must estimate the slicing uncertainty due to our 
	cutoff of $q_T^\text{cut}=\SI{5}{\GeV}$.
	With a \SI{5}{\GeV} slicing cutoff, and including linear power corrections, we obtain a 
	value of $-22.6\pm \SI{1.4}{pb}\, (\text{numerical})$. Using a \SI{10}{\GeV} slicing cutoff 
	instead we obtain a value of $-21.3\pm \SI{1.4}{pb} \,
	(\text{numerical})$. From this variation we therefore assign a further slicing uncertainty of 
	about \SI{1}{pb}.
	
	\begin{figure}
		\includegraphics[width=\columnwidth]{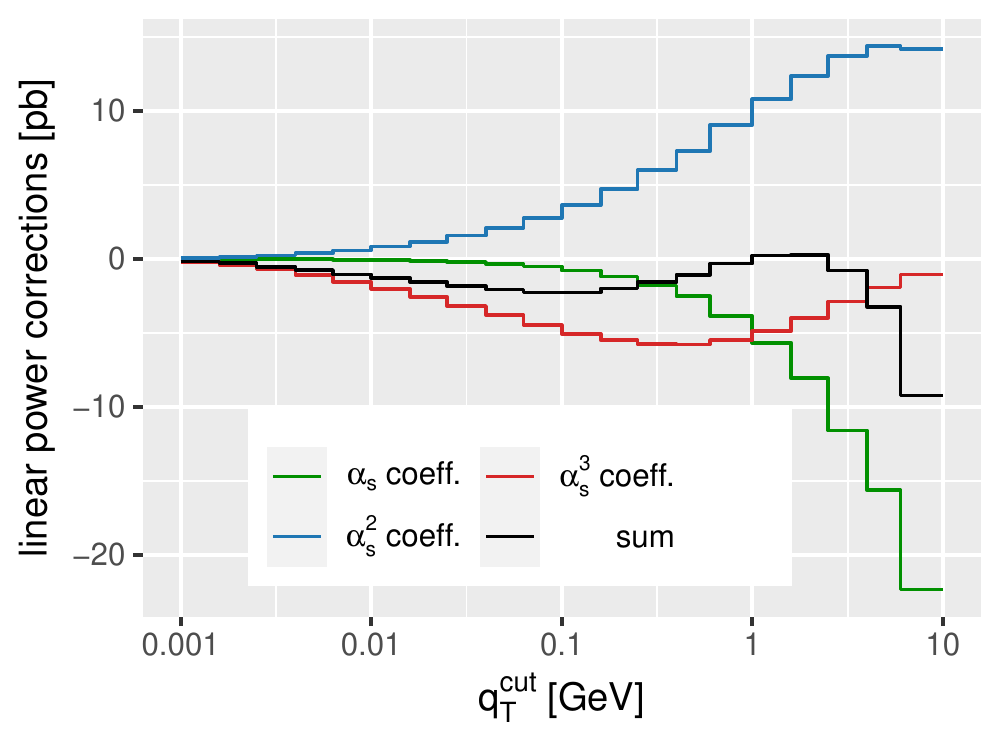}
		\caption{Linear power corrections as a function of the $q_T$ slicing cutoff for the 
		fiducial 
			cross-section with \CMS{} cuts as in the 
			results section. Of most significance are the \NNNLO{} 
			coefficient power corrections, since no local subtraction method exists at this order 
			that 
			would avoid slicing power corrections.}
		\label{fig:linpc}
	\end{figure}
	
	This uncertainty estimate is also supported by an examination of the linear power corrections,
	which can be easily computed in our formalism.
	The knowledge of the size of the linear power corrections allows another estimate of 
	slicing cutoff effects, since further corrections should be even more power suppressed.
	\Cref{fig:linpc} shows the size of the linear power corrections for our \CMS{} cuts in 
	the results section.
	For the \NNNLO{} 
	coefficient the linear power corrections are accidentally small around a slicing cutoff of 
	\SI{10}{\GeV} and would be larger for a \SI{1}{\GeV} slicing cutoff. Around \SI{0.5}{\GeV} the 
	linear power corrections at \NNNLO{} are largest, about $\SI{-5}{pb}$. Knowing that further 
	slicing 
	cutoff power corrections are suppressed by a factor of $q_T^\text{cut}/Q$, we expect further 
	corrections of less than \SI{1}{pb} with a cutoff of \SI{10}{\GeV} or less. This is in line 
	with 
	our cutoff variation discussed above. Ultimately in the presence of symmetric cuts a slicing 
	uncertainty of about \SI{1}{pb} is unavoidable for slicing calculations and can only be cured 
	by a local 
	subtraction scheme at fixed-order.
	
	The final comparison with the {\abbrev NNLOjet} result of ref.~\cite{Chen:2022cgv}, for
	the \NNNLO{} corrections is:
	\begin{gather}
		\text{this work: } \SI{-22.6}{pb} \pm \SI{1.4}{pb}\, (\text{num.})
		\pm \SI{1}{pb}\, (\text{slicing}) \nonumber \\
		\text{Ref.~\cite{Chen:2022cgv}: } \SI{-18.7}{pb} \pm \SI{1.1}{pb} \,
		(\text{num.}) \pm \SI{0.9}{pb} \,(\text{slicing}) \nonumber
	\end{gather}
	The {\abbrev NNLOjet} result is obtained with a slicing 
	cutoff of $q_T^\text{cut}=\SI{0.8}{\GeV}$ that is varied by a factor of two to obtain the 
	slicing 
	uncertainty. Our estimated slicing uncertainty is similar, which is only because the (dominant) 
	linear power corrections are known. Cuts that eliminate the linear power corrections altogether 
	\cite{Salam:2021tbm} improve this situation \cite{Chen:2022cgv}.
	
	The two results are in agreement within 
	mutually large uncertainties of about 10\% on the \NNNLO{} coefficient. Fortunately this 
	uncertainty on the \NNNLO{} \emph{coefficient} reduces to about three to four per mille for the 
	full result and is currently insignificant compared to truncation uncertainties and the 
	experimental precision (which is limited by a 2\% luminosity
	uncertainty). But in the future local subtraction 
	methods for \NNNLO{} are clearly preferred.

	\section{Results}
	\label{sec:results}
	
	We present results at $\sqrt{s}=\SI{13}{\TeV}$ using the \texttt{NNPDF4.0} \PDF{} set at 
	\NNLO{} with $\alpha_s(m_Z)=0.118$ \cite{NNPDF:2021njg}. Electroweak input parameters are 
	chosen in the $G_\mu$ scheme with 
	$m_Z=\SI{91.1876}{\GeV}$, $m_W=\SI{80.385}{\GeV}$, $\Gamma_Z=\SI{2.4952}{\GeV}$ and
	$G_F=\SI{1.16639e-5}{\GeV^{-2}}$.
	We denote the matched resummation accuracy with 
$\alpha_s$ for \NNLL{}+\NLO{}, $\alpha_s^2$ for \NNNLL{}+\NNLO{} and $\alpha_s^3$ for 
\NFLLp{}+\NNNLO{}.

Our fiducial selection cuts in \cref{tab:zcms13tev} 
are chosen to compare with the most recent $Z$-boson precision measurement by \CMS{} in 
ref.~\cite{CMS:2019raw}.
The 
symmetric lepton cuts used in this analysis cause a poor 
perturbative convergence for fixed-order calculations and can also lead to numerical issues.
However, the use of resummation resolves such
issues \cite{Ebert:2020dfc,Salam:2021tbm,Billis:2021ecs}.
	
	 In our calculation we distinguish between three scales for 
	estimating uncertainties. We use a low (resummation) scale $\sim q_T$ (see 
	ref.~\cite{Becher:2020ugp} for 
	details) to which \RGE{}s are evolved down from 
	the hard scale chosen as 
	$\sqrt{m_Z^2+p_{T,Z}^2}$. The \CuTeMCFM{} resummation formalism 
	\cite{Becher:2010tm,Becher:2011xn,Becher:2012yn} is originally derived using an analytic 
	regulator to regulate rapidity divergences in the transverse 
	position dependent \PDF{}s (collinear anomaly formalism). This is opposed to using a rapidity 
	regulator 
	that introduces a rapidity scale 
	\cite{Chiu:2012ir}. We 
	have re-introduced a scale estimating the effect of a different rapidity scale as suggested in 
	ref.~\cite{Jaiswal:2015nka}. We vary hard and low scale by a factor of two, and rapidity scale 
	by a factor of six, tuned on the truncation of the improved power counting, to obtain a robust 
	estimate of truncation uncertainties. Most importantly 
	our formalism allows for the variation 
	of the low scale, which dominates uncertainties at small $q_T$. Last, in our uncertainty bands 
	we include the effect of varying the transition function in the region of about 
	\SIrange{40}{60}{\GeV} where matching corrections become significant, following the same 
	procedure as in ref.~\cite{Becher:2020ugp} at a lower order.
	
While for Drell-Yan production our resummation formalism does not set the central low scale 
below $\sim \SI{2}{\GeV}$ \cite{Becher:2020ugp}, a downwards variation would probe close towards 
the non-perturbative regime. We therefore set a minimum of \SI{2}{\GeV} and symmetrize the 
uncertainty bands since the variation becomes ineffective at small scales. Note that about $2\%$ of 
the total fiducial cross-section comes from the region $q_T < \SI{1}{\GeV}$ where one might expect 
additional non-perturbative effects of an unknown size.

	The \CMS{} collaboration~\cite{CMS:2019raw} provides both differential results to compare with 
	as well as a total fiducial 
	cross-section measurement, that we discuss in turn below.

	\begin{table}
		\centering
		\caption{Fiducial cuts for $Z\to l^+ l^-$ 
		used in the \CMS{} \SI{13}{\TeV} analysis \cite{CMS:2019raw}.}
		\vspace*{0.5em}
		\bgroup
		\setlength\tabcolsep{1em}
		\def\arraystretch{1.5}%
		\begin{tabular}{l|c}
			Lepton cuts & $q_T^{l} > \SI{25}{\GeV}, \abs{\eta^l} < 2.4$\\
			Separation cuts  & $\SI{76.2}{GeV} < m^{l^+l^-} < \SI{106.2}{GeV}$, \\
			 &  $\abs{y^{l^+l^-}} < 2.4$
		\end{tabular}
		\egroup
		\label{tab:zcms13tev}
	\end{table}

	\subsection{Differential results}

\begin{figure}[t]
	\includegraphics[width=\columnwidth]{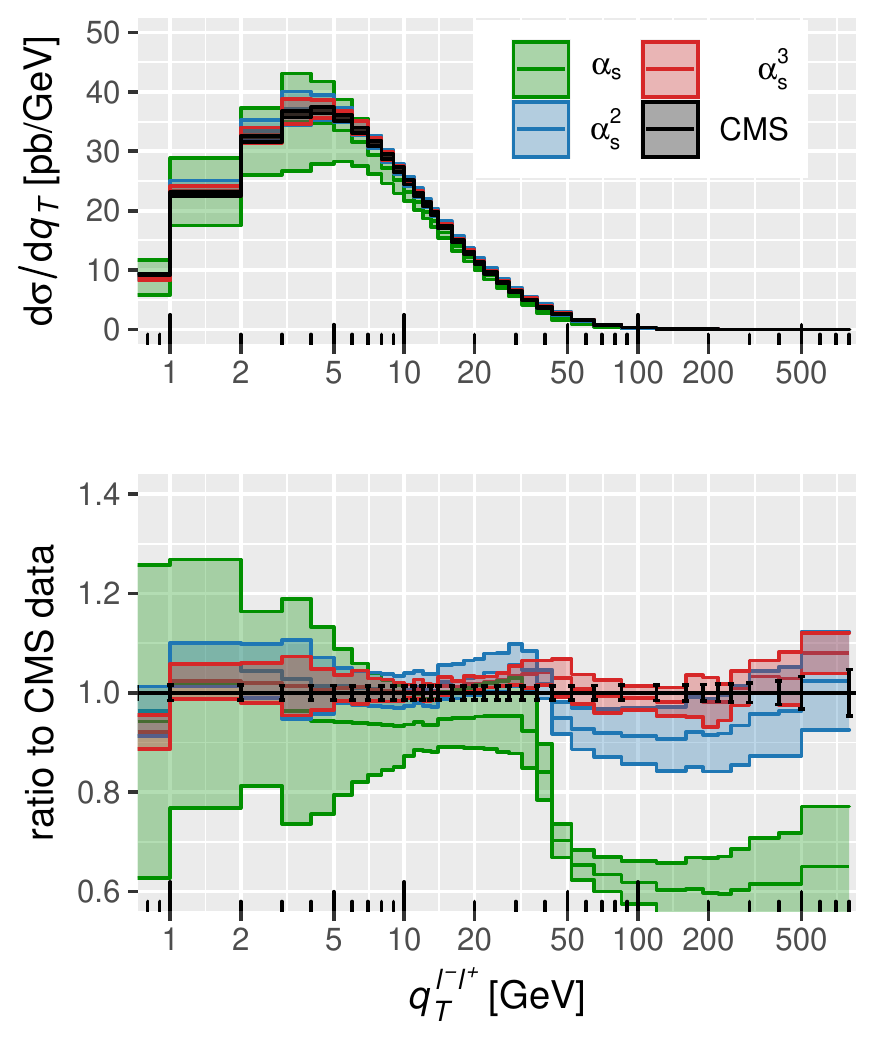}
	\caption{Differential transverse-momentum resummation improved predictions for the 
		$q_T^{l^- l^+}$ distribution at order $\alpha_s$, $\alpha_s^2$ and $\alpha_s^3$. }
	\label{fig:qt}
\end{figure}
	
	In \cref{fig:qt} we present the $Z$ boson transverse momentum distribution predictions at order 
	$\alpha_s$, $\alpha_s^2$ and $\alpha_s^3$ and 
	compare it to the \CMS{} \SI{13}{\TeV} measurement \cite{CMS:2019raw} with cuts as in  
	\cref{tab:zcms13tev}.
	
	 Overall there is an excellent agreement between theory and data at the highest order.
	 Going 
	 from $\alpha_s^2$ to $\alpha_s^3$ decreases uncertainties and improves agreement with data 
	 noticeably at both large and small $q_T$. In the first 
	 bin $\SI{0}{\GeV} < q_T < \SI{1}{\GeV}$ we notice a relatively large difference to the data, 
	 but this is also where one would expect a non-negligible contribution from non-perturbative 
	 effects.
	 We note that the impact of the corrections 
	included in \NFLLp{} is a noticeable shift in this distribution,
	compared to \NNNLLp{}, as discussed further in
	\cref{sec:nthreenfourll}.
		
	\begin{figure}[t]
		\includegraphics[width=\columnwidth]{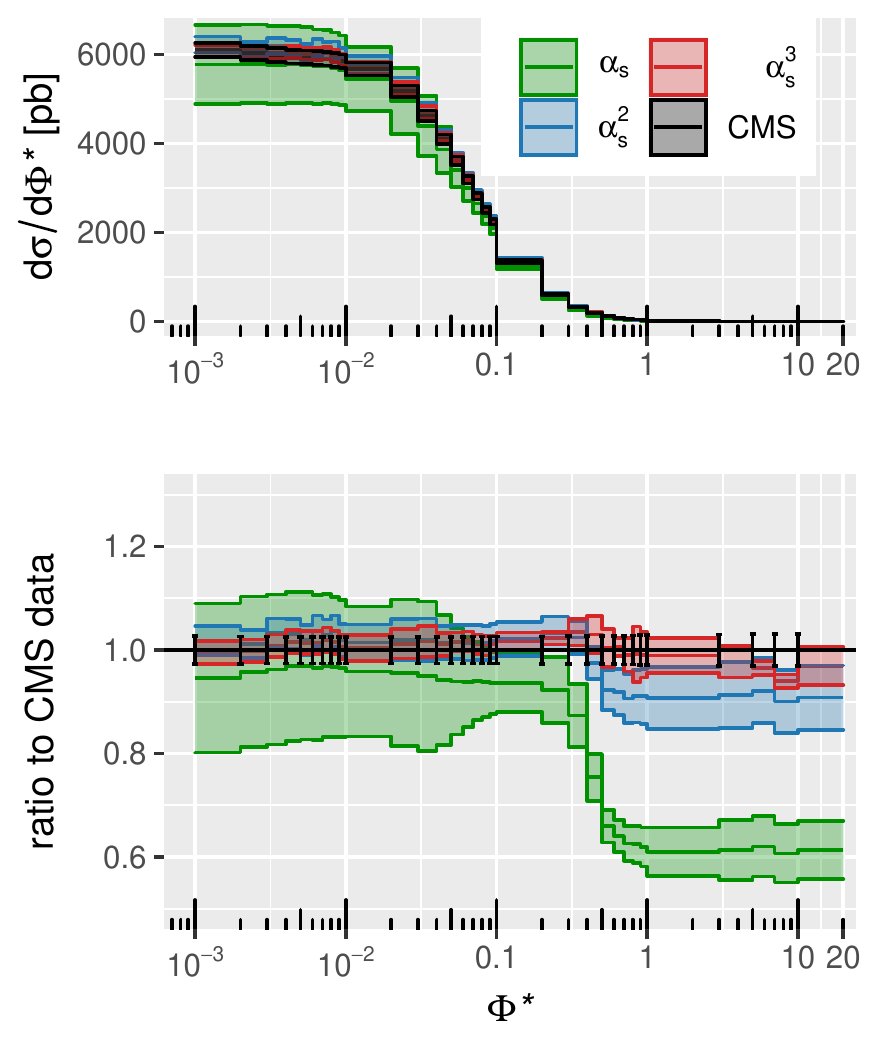}
		\caption{Differential transverse-momentum resummation improved predictions for the 
			$\Phi^*$ distribution at order $\alpha_s$, $\alpha_s^2$ and $\alpha_s^3$. }
		\label{fig:phistar}
	\end{figure}

	For the $\Phi^*$ distribution shown in \cref{fig:phistar} results are overall very similar. For 
	the transverse momentum 
	distribution we neglect matching corrections at $\alpha_s^3$ below $q_T<\SI{5}{\GeV}$. Here we 
	correspondingly	neglect them below $\Phi^* < \SI{5}{\GeV}/m_Z \sim 0.05$ and at lower orders 
	below $\Phi^* < \SI{1}{\GeV}/m_Z \sim 0.01$, an overall per-mille level effect in that region.
	
\begin{figure}[t]
	\includegraphics[width=\columnwidth]{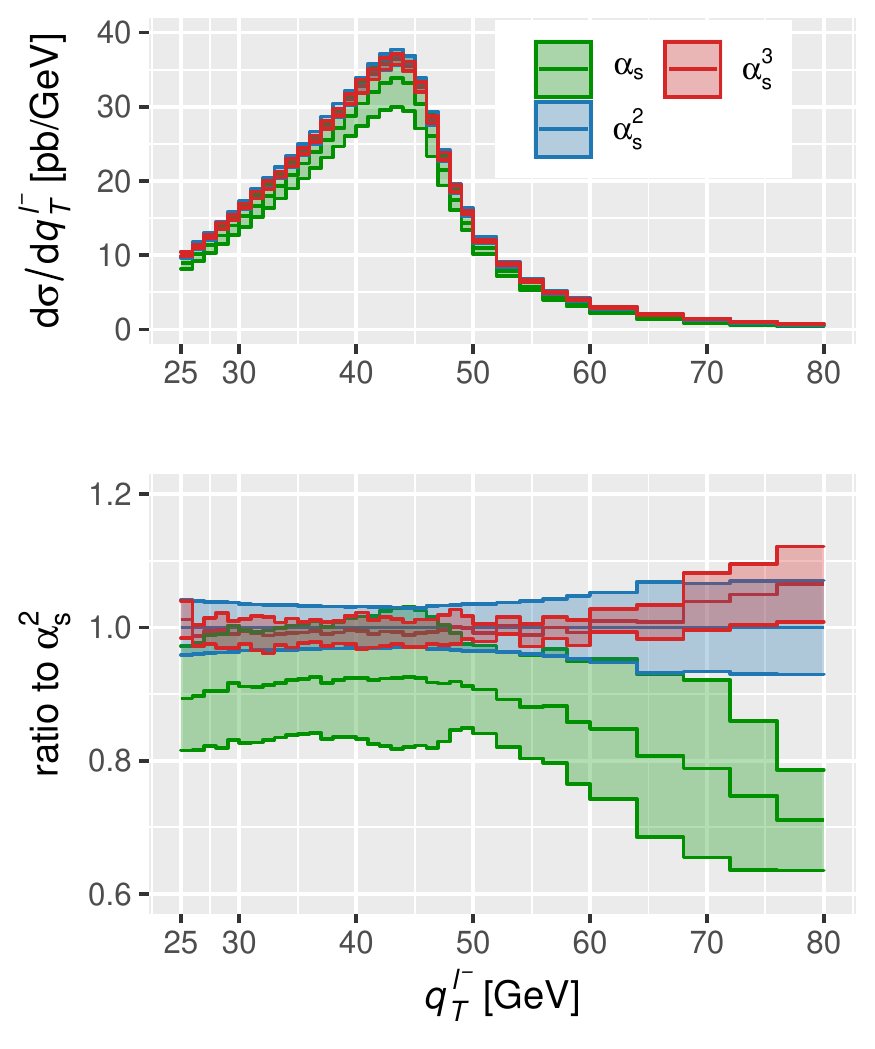}
	\caption{Differential transverse-momentum resummation improved predictions for the 
		lepton transverse momentum distribution at order $\alpha_s$, $\alpha_s^2$ and $\alpha_s^3$. 
		}
	\label{fig:pt3}
\end{figure}

	Since our resummation implementation is fully differential in the electroweak final state we 
	can naturally also present the transverse momentum distribution of the final state lepton, see 
	\cref{fig:pt3}. This is plagued by a Jacobian peak at fixed-order and crucially requires 
	resummation. The higher-order $\alpha_s^3$ corrections further stabilize the results with 
	smaller uncertainties.
	
	\subsection{Total fiducial cross-section}
	
	\begin{table*}
		\centering
		\caption{Fiducial cross-sections in pb for the cuts in \cref{tab:zcms13tev} and input 
		parameters as in the text. Uncertainties for the resummation-improved results include  
		matching to fixed-order (mat.), neglected matching corrections (m.c.), and
		by scale variation (sc.).  The fixed-order
		result at \NNNLO{} has an additional slicing-cutoff uncertainty.
		For comparison, the final row shows the CMS measurement
		(for electron and muon channels combined) \cite{CMS:2019raw}. }
		\vspace*{0.5em}
		\bgroup
		\setlength\tabcolsep{1em}
		\def\arraystretch{1.5}%
		\begin{tabular}{c|c|c}
			Order $k$ & fixed-order $\alpha_s^k$ & res. improved $\alpha_s^k$ \\
			0 & $694^{+85}_{-92}$ & --- \\
			1 & $732^{+19}_{-30}$ & $637\pm8_\text{mat.} \pm 70_\text{sc.}$ \\
			2 & $720^{+4}_{-3}$ & $707\pm3_\text{mat.}\pm 29_\text{sc.}$ \\
			3 & $700^{+4}_{-6}\pm 1_\text{slicing}$ & $702\pm1_\text{mat.}  \pm 
			1_\text{m.c.} \pm 17_\text{sc.}$  \\
		\hline
		\multicolumn{3}{c}{$699\pm 5\, (\text{syst.}) \pm 17\, (\text{lumi.})$ ($e$, $\mu$ combined) \cite{CMS:2019raw}}\\
		\end{tabular}
		\egroup
		\label{tab:zcms13tev_fiducialcross}
	\end{table*}
	
	In \cref{tab:zcms13tev_fiducialcross} we present total fiducial cross sections. Uncertainties 
	of the fixed-order \NNLO{} ($\alpha_s^2$) result, obtained by taking the envelope of a variation of renormalization and factorization 
	scales 	by a factor of two, are particularly small at the level of 
	$0.5\%$ and do not improve towards \NNNLO{} with large corrections. The resummation improved 
	results are obtained by integrating over 
	the matched $q_T$ spectrum shown in \cref{fig:qt}. Uncertainties of the resummation 
	improved 
	predictions are obtained by taking the envelope of the variation of hard, low and rapidity 
	scales in the fixed-order and resummation region.
	The matching uncertainty from the transition function variation is quoted separately.
	We estimate the effect of neglecting matching corrections at $\alpha_s^3$ below 
	$q_T\leq \SI{5}{\GeV}$ to be less than \SI{1}{pb}.
	
	The resummation improved result at 
	$\alpha_s$ has large uncertainties that stem from an insufficient order of the 
	resummation (\NNLL{}), which still has substantial uncertainties in the Sudakov peak region
	(c.f. \cref{fig:qt}). The results quickly stabilize, with less than a percent difference between
        the central $\alpha_s^2$ and $\alpha_s^3$ predictions. Nevertheless, 
	the uncertainties we obtain are noticeably larger than the fixed-order uncertainties.
	We further observe that going from \NNNLL{}/$\alpha_s^2$ to \NFLLp{}/$\alpha_s^3$ does not
	reduce uncertainties as substantially as when going from $\alpha_s$ to $\alpha_s^2$. This is 
	because the resummation uncertainties around the Sudakov peak region at small 
	$q_T\sim\SI{5}{\GeV}$ do not improve dramatically.
	
	While this behavior, of only moderately decreasing uncertainties going from $\alpha_s^2$ to 
	$\alpha_s^3$, is consistent with the findings of ref.~\cite{Chen:2022cgv} using {\abbrev 
	RadISH} 
	resummation, our uncertainties of the resummation improved fiducial cross-section are larger 
	than the uncertainties presented there. Our $\alpha_s^3$ prediction has 
	uncertainties of about $2.5\%$, while using {\abbrev RadISH} for the resummation results in 
	uncertainties 
	of about $1\%$. Given that differentially in \cref{fig:qt} we see still some variation in the 
	low $q_T$ region between the 
	central $\alpha_s^2$ and $\alpha_s^3$ results, we are confident in our more conservative 
	uncertainty estimate.
	
	Indeed, theory uncertainties have become an important topic within recent years 
	\cite{Duhr:2021mfd}. First, they cannot be interpreted statistically and second, perturbative 
	predictions are limited to the level presented here for the foreseeable future. It is therefore 
	important to study them with as much scrutiny as possible.
	An approach followed in ref.~\cite{Camarda:2021ict} has been to take half the difference 
	between the two highest order results as an uncertainty. This would bring our uncertainties 
	closer in line with the uncertainties presented in ref.~\cite{Chen:2022cgv}, less than one 
	percent.

	\section{Conclusions \& Outlook.}
	\label{sec:conclusions}
	
	$Z$-boson production is the most precisely measured process at the \LHC{} and meanwhile solely 
	limited in precision by the beam luminosity uncertainty. At the same time it is one of the most 
	important standard candles and enters many precision prediction ingredients like \PDF{}s 
	and \SM{} input parameters. It is crucial that theory predictions are available at the 
	same level of precision to make best use of the available measurements.
	
	In this paper we presented the first transverse-momentum ($q_T$) resummation improved 
	calculation at 
	the level of \NFLLp{}+\NNNLO{}, which broadly reduces theory uncertainties to the few percent 
	level. Our results show excellent agreement with the \SI{13}{\TeV} \CMS{} measurements within a 
	few percent both at the differential level from $q_T^Z=\SI{1}{\GeV}$ to $\sim$\SI{500}{\GeV} 
	and 
	for $\Phi^*$ over the whole spectrum, as well as for the total fiducial cross-section. As a 
	consequence of the resummation (and inclusion of linear power corrections), our 
	calculation can provide reliable predictions also for past experimental analyses
	that would induce factorially divergent contributions at fixed order due to cuts, e.g. 
	symmetric lepton cuts 
	\cite{Salam:2021tbm}.
	
	All previous calculations of order \NNNLL{}$^\prime$+\NNNLO{} rely on a single 
	$Z$+jet \NNLO{} calculation \cite{Gehrmann-DeRidder:2015wbt}. Further, uncertainties (via scale 
	variation) for resummation improved results were only estimated by using the {\abbrev RadISH} 
	resummation framework \cite{Re:2021con,Bizon:2017rah}. Due to the utmost importance of this 
	process, it is crucial to provide an independent calculation using completely different 
	methods to reliably estimate uncertainties. It allows future (experimental) studies to assess 
	the validity of their input theory predictions through independent results.
	This becomes 
	increasingly important with the advent of very precise collider measurements
	that might indicate tension with the \SM{} \cite{CDF:2022hxs}.
	The public availability of our calculation as part of the upcoming \CuTeMCFM{} release allows 
	for a much larger audience to make use of this state-of-the-art precision, to implement 
	modification of cuts and input parameters, and also to re-use parts and to validate other 
	calculations \cite{Caola:2022ayt}.
	
	Previously it was found that fiducial cross-section uncertainties at the level 
	of $\alpha_s^3$ are similar to those at $\alpha_s^2$, about $1\%$ using {\abbrev RadISH} 
	resummation \cite{Chen:2022cgv}. With resummation, this uncertainty is 
	dominated by the uncertainties around the Sudakov peak at small $q_T$, i.e. mostly within the 
	pure resummation region. We find more conservative uncertainties of about 
	$2.5\%$ using \CuTeMCFM{} resummation.
	
        Although the theoretical precision of the calculation discussed in this paper is now
        at an impressive level, there are two important
	aspects that require further work.  Statistical 
	\PDF{} uncertainties have reached the level of one percent 
	\cite{NNPDF:2021njg,Bailey:2020ooq} and systematic effects can no longer be neglected. Since 
	these 
	uncertainties are at the same level as perturbative truncation uncertainties, a careful study 
	of \PDF{} effects at this order will be an important future direction. 
	Indeed, while finalizing this manuscript, approximate \NNNLO{} \PDF{}s have been 
	introduced by 
	the {\abbrev MSHT} group \cite{McGowan:2022nag}. They take into 
	account approximations for the four loop splitting functions through known information on small 
	and large $x$ and available Mellin moments. Such theory approximations of missing higher-order 
	effects are included in their Hessian procedure as nuisance parameters.\footnote{
	A preliminary study of the potential impact of this \PDF{} set on the results shown in this
	paper is presented in \cref{appendix}.}

	In addition, in order to better match with data at very small $q_T$, it is  possible to
	include a parametrization of non-perturbative effects,
	see e.g. refs.~\cite{Becher:2013iya,Ebert:2022cku}. 
	This can then inform the modeling of the related process of
	$W$-boson production and thus have implications for the extraction of the $W$-boson mass.
	Extending $W$-boson production in \CuTeMCFM{} to $\alpha_s^3$ accuracy will thus be a 
	valuable extension that allows for very precise $W/Z$ boson ratio predictions \cite{Ju:2021lah}.

	\paragraph{Acknowledgments.} We would like to thank Thomas Becher and Robert Szafron for 
	helpful discussions. 
	Furthermore, we 
	would like to thank {\abbrev NERSC} for use of the Perlmutter 
	supercomputer that enabled this calculation.
        This manuscript has been authored by Fermi Research Alliance, LLC under Contract No.
        DE-AC02-07CH11359 with the U.S. Department of Energy, Office of Science, Office of High Energy Physics.
	Tobias Neumann is supported by 
	the United States Department of Energy under Grant Contract DE-SC0012704.
	This research used 
	resources of the National Energy Research Scientific Computing Center, which is supported by 
	the Office of Science of the U.S. Department of Energy under Contract No. DE-AC02-05CH11231. 
	Some of the numerical calculations entering the results in this paper were performed using
	the Wilson High-Performance Computing Facility at Fermilab.
	
	\appendix

	\section{Impact of \NNNLO{} \PDF{}s.}
	\label{appendix}
	
\begin{figure}
		\includegraphics[width=\columnwidth]{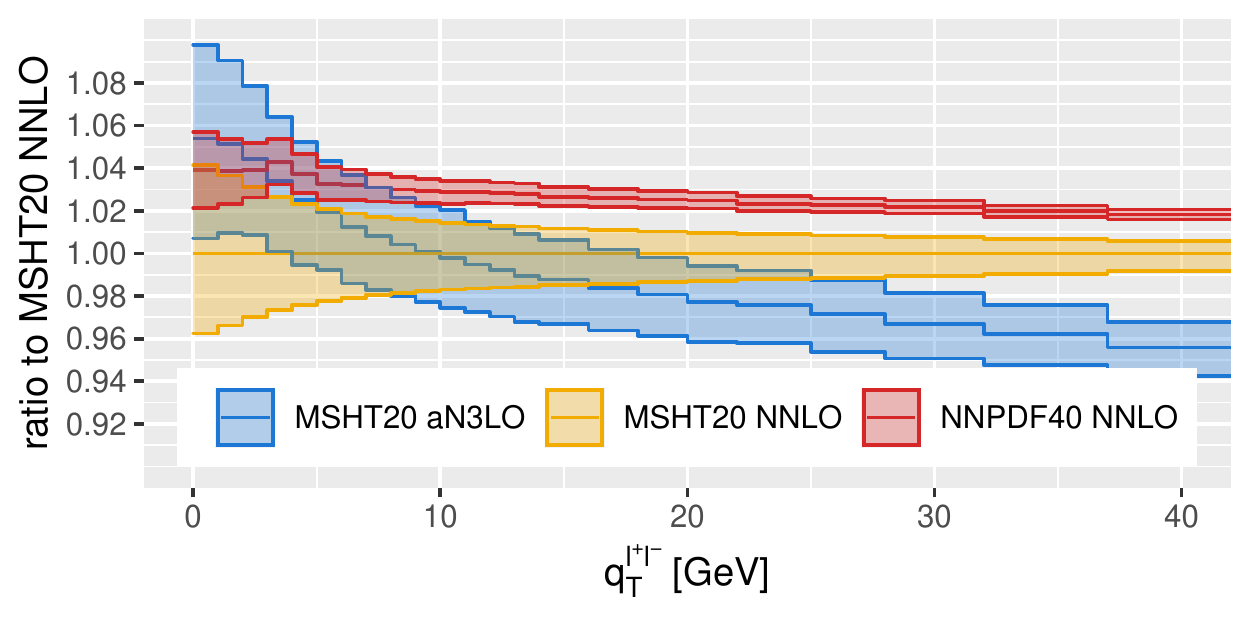}
		\caption{\PDF{} uncertainties of the purely resummed {\abbrev N$^4$LL} $q_T$ spectrum as 
		the ratio to 
		the 
		{\abbrev MSHT20 NNLO} central value.}
		\label{fig:n3lopdfs}
	\end{figure}

	Here we give a first 
	impression of the impact of the approximate \NNNLO{} \PDF{}s of Ref. \cite{McGowan:2022nag} by
	comparing the \PDF{} uncertainties of this set to our default set 
	{\abbrev NNPDF40 NNLO} \cite{NNPDF:2021njg} and to {\abbrev MSHT20 NNLO} \cite{Bailey:2020ooq}.
	\Cref{fig:n3lopdfs} shows the purely 
	resummed spectrum up to 
	\SI{40}{\GeV}, where matching corrections of about 5\% are neglected at \SI{20}{\GeV} (less 
	than 2\% below \SI{10}{\GeV}). We do not expect that the matching corrections change the 
	relative \PDF{} results 
	and uncertainties substantially. About two-thirds of the total fiducial cross-section 
	originates from the 
	integrated purely resummed spectrum up to \SI{20}{\GeV}. The results demonstrate that 
	systematic differences between \PDF{} sets are still dominant, comparable to the effect of 
	\NNNLO{} corrections in the \PDF{}s. Uncertainties for the {\abbrev MSHT20 a\NNNLO{}} \PDF{} 
	set are larger since it includes missing higher-order effects with the \PDF{} uncertainties. 
	Overall, combined statistical and systematic \PDF{} uncertainties are comparable to the residual 
	truncation uncertainties found in our paper.
	
\section{Comparison of \NNNLLp{} with \NFLLp{}.}
\label{sec:nthreenfourll}

\begin{figure}[h]
	\includegraphics[width=\columnwidth]{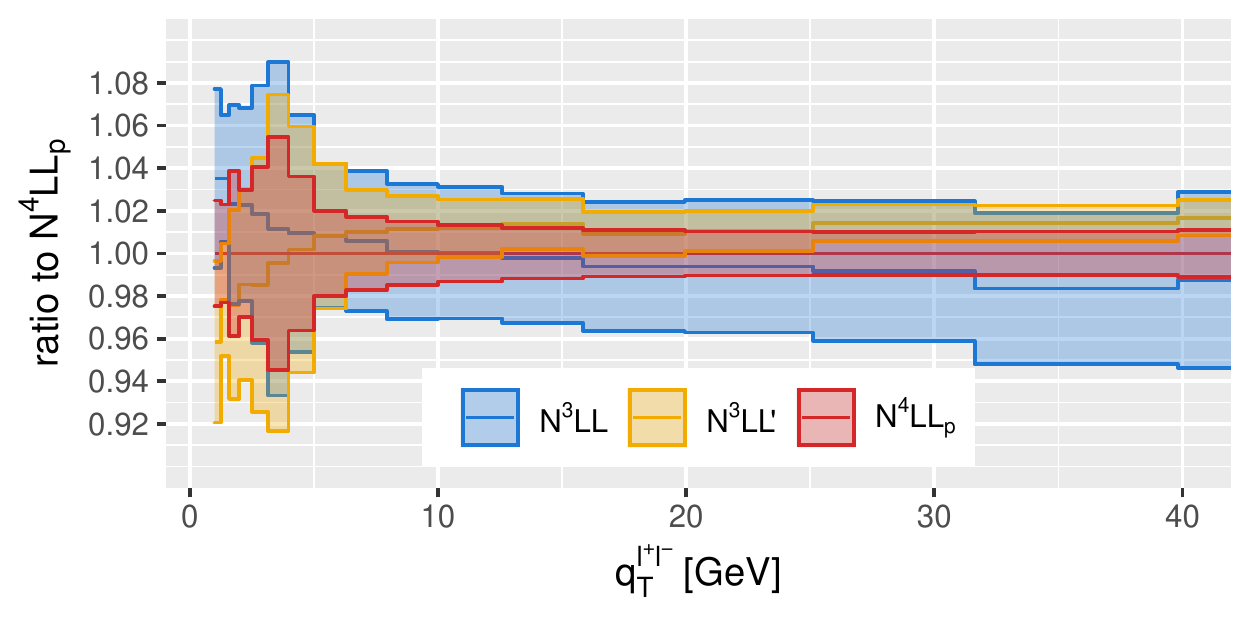}
	\caption{Purely resummed $q_T$ spectrum at different logarithmic orders as the ratio to 
		\NFLL{} using {\abbrev NNPDF40 NNLO} \PDF{}s in all cases.}
	\label{fig:resumorders}
\end{figure}

In fig.~\ref{fig:resumorders} we show the purely resummed $q_T$ spectrum at order \NNNLL{}, 
\NNNLLp{} and \NFLL{} normalized to \NFLL{} and using {\abbrev NNPDF40 
	NNLO} \PDF{}s in all cases. The \NNNLLp{} result does not include the four-loop rapidity 
anomalous dimension 
(the additional contribution from the estimated five-loop cusp anomalous dimension to the 
hard function evolution is completely negligible).
This figure indicates that a substantial decrease in uncertainties 
comes from the \NNNLLp{} result, with little additional reduction at \NFLL{}.
However the \NFLL{} result shifts noticeably, and 
its central value is only marginally compatible with the \NNNLLp{} uncertainty estimate. It is 
therefore an important step in the full $\alpha_s^3$ precision. 

	\bibliographystyle{JHEP}
	\bibliography{refs}
	
\end{document}